\documentclass{article}

\usepackage[numbers]{natbib}
\usepackage{rotating}
\usepackage{graphicx}
\usepackage{bm}
\usepackage[margin=0.5in]{geometry}
\title{On Direction Preserving Discretizations for Computing Phase-Space Densities}

\author{David J. Chappell$^{(a)}$, Martin Richter$^{(a,b)}$ and Gregor Tanner$^{(b)}$\vspace{2mm}\\
$^{(a)}$ School of Science and Technology, \\Nottingham Trent
University, \\Clifton Campus, Clifton Lane, \\ Nottingham NG11 8NS, UK\vspace{2mm}\\
$^{(b)}$ School of Mathematical Sciences, \\University of
Nottingham,\\ University Park, \\Nottingham NG7 2RD, UK}
\date{}
\begin{document}
\maketitle
\begin{abstract}
Ray flow methods provide efficient tools for modelling wave energy transport in complex systems at high-frequencies.  We compare two Petrov-Galerkin discretizations of a phase-space boundary integral model for stationary wave energy densities in two-dimensional domains.  The directional dependence is approximated using a finite set of directions oriented into the domain from the boundary.  The propagation direction can be preserved across multi-component domains when the directions within the local set for a given region of the boundary are taken as a subset of a global direction set.  In this work we compare the use of piecewise constant and piecewise linear test functions,  which physically corresponds to the interpolation scheme used when the transport is in a direction not belonging to the finite global set.
\end{abstract}

% Head 1
\section{INTRODUCTION}
Dynamical energy analysis (DEA) is an approach for modelling wave energy densities at high-frequencies that was first proposed just over ten years ago \cite{CTLS13, TH19, GT09}.  DEA is based on a linear integral operator description of phase-space density transport along ray trajectories between positions on the boundary of a domain or sub-domain.  Recent developments have seen the capability of DEA extended to industrial applications  \cite{CTLS13, TH19},  as well as stochastic propagation through uncertain structures \cite{Chaos14,  Uncert2,  Uncert3}.

A Petrov-Galerkin method that is efficient for modelling densities with strong directional dependence was recently proposed \cite{DeltaDEA}.  This class of problems has usually proved problematic for the DEA method \cite{CTLS13}.  In \cite{DeltaDEA}, a basis approximation using Dirac delta distributions was proposed to approximate the directional dependence. The direction of transport can be preserved throughout multi-domains provided that the Dirac delta specified direction set local to any part of the boundary is inherited from a common global set of directions; the propagation is then able to continue along rays with the same global direction through different sub-domains.  A beneficial consequence is that the proposed methodology can be applied directly on complicated domains formed from a potentially large number of simpler sub-domains as is typically the case with the finite element type meshes used for industrial applications.  

In this study we investigate the effect of modifying the choice of test functions from the piecewise constant indicator functions employed in the Petrov-Galerkin scheme proposed in \cite{DeltaDEA}.  For classical potential problems, it is known that the convergence of  Petrov-Galerkin schemes based on Dirac delta distribution basis approximations together with splines as test functions crucially depends on this choice \cite{Dirac1, Dirac2, Dirac3}. In particular,  higher order test functions were shown to lower the regularity requirements on the boundary data.  In comparison to more standard projection methods for integral equations,  such as the collocation and Galerkin methods,  these Petrov-Galerkin schemes combine the efficiency and ease of implementation of collocation based schemes with potentially even lower regularity requirements for convergence than the Galerkin method \cite{Dirac2}.  These features are important in this work since ray tracing solutions can often have low regularity and the added dimensionality of working in phase-space means that a faster implementation with fewer integrals to compute is desirable. 

\section{PHASE-SPACE BOUNDARY INTEGRAL MODEL}\label{sec:PropDenOp}
%To here
We are concerned with transporting densities along ray trajectories through multi-domains $\Omega=\cup_{j=1}^{K}\Omega_j$. The ray dynamics in $\Omega_j$ is defined by the Hamiltonian $H_j(\mathbf{r},\mathbf{p})=|\mathbf{p}|/\eta(\mathbf{r})\equiv1$ for $j=1,\ldots,K$.  The phase-space coordinates $(\mathbf{r},\mathbf{p})$ denote the position $\mathbf{r}$ and momentum $\mathbf{p}$ vectors,  respectively.  Each $\Omega_j$, $j=1,\ldots,K$ is assumed to be a convex polygon containing a homogeneous medium.  As a consequence we may write $\eta(\mathbf{r})=\eta_j$ when $\mathbf{r}\in\Omega_j$ for $j=1,\ldots,K$,  where $\eta_j$, $j=1,\ldots,K$ are constants.  When $\eta_j=c_j^{-1}$ is taken to be the inverse of the phase speed in $\Omega_j$,  then $H_j$ defines the ray trajectories obtained via leading order high-frequency asymptotics for the Helmholtz equation
\begin{equation}\label{eq:Helm}
c_j^2\:\Delta u + \omega^2 u =0,
\end{equation}
at angular frequency $\omega$.  The choice of $\eta$ can easily be modified for flexural waves with nonlinear dispersion - see \cite{DeltaDEA}.

\begin{figure}[h]
 \centerline{\includegraphics[trim=0 50 0 0, width=150pt]{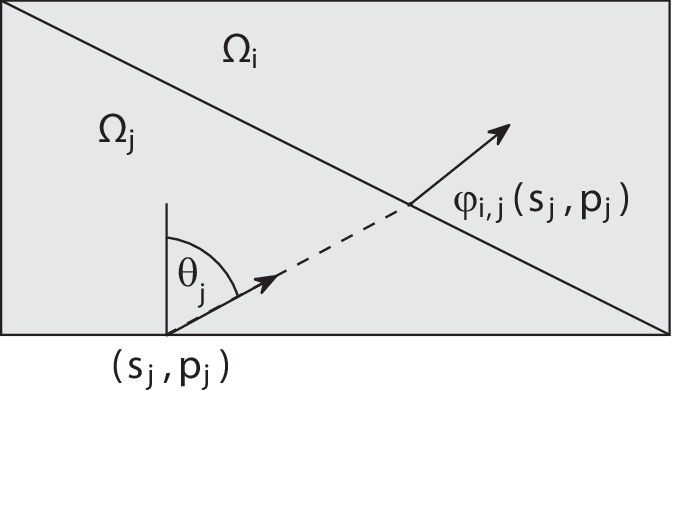}}
\caption{The boundary mapping $\varphi_{i,j}(s_j,p_j)=(s'_i,p'_i)\in\Gamma_i\times(-\eta_i,\eta_i)$ of the
phase-space boundary coordinates $X_j=(s_j,p_j)\in\Gamma_j\times(-\eta_j,\eta_j)$, which maps to the next intersection with a boundary edge followed by either a specular reflection or transmission.}\label{fig:BMap}
\end{figure}
In order to express our model in boundary integral form it is convenient to introduce the phase-space boundary coordinates $Y_j=(s_j,p_j)$ on $\Gamma_j=\partial\Omega_j$, $j=1,\ldots,K$,  where $s_j$ is an arclength parameter for $\Gamma_j$ and $p_j=\eta_j\sin(\theta_j)$ is the tangential component of $\mathbf{p}$ at $s_j$.  Here $\theta_j$ is the angle formed between the ray (oriented into $\Omega_j$) and the inward normal vector at $s_j$ as shown in Fig.~\ref{fig:BMap}.  We next introduce a local boundary flow map $\varphi_{i,j}(s_j,p_j)=(s'_i(s_j,p_j),p'_i(s_j,p_j))$,  which describes the discrete evolution of the rays at times coinciding with boundary intersections.  As written here,  $\varphi_{i,j}$ maps the boundary phase-space coordinate $(s_j,p_j)$ on the boundary of $\Omega_j$ to $(s'_i(s_j,p_j),p'_i(s_j,p_j))$  on the boundary of $\Omega_i$. To write  $\varphi_{i,j}$ in this form we have implicitly assumed that either $i=j$, or $\Gamma_i$ and $\Gamma_j$ share a common edge through which the ray can transmit.  As before,  $p'_i(s_j,p_j)=\eta_i\sin(\theta'_i(s_j,p_j))$ denotes the tangential slowness at $s_i'$,  and $\theta'_i$ is the angle formed between the outgoing ray and the inward normal vector to $\Gamma_i$ at $s'_i$.  In the simplest cases,  $\theta'_i$ is obtained from either a specular reflection when $i=j$, or if $i\neq j$ and $\eta_i=\eta_j$ then the ray will continue in the same direction into $\Omega_i$.

Phase-space densities are transported throughout $\Omega$ using a local boundary operator $\mathcal{B}_{j}$,  which transports a density $f$ along the boundary flow $\varphi_{i,j}$ as follows \cite{CTLS13, DeltaDEA}
\begin{equation}\label{eq:OpB}
\mathcal{B}_{j}[f](X_i):= \int e^{-\mu_j D(X_i,Y_j)}w_{i,j}(Y_j) \delta(X_i-\varphi_{i,j}(Y_j)) f(Y_j)
\, \mathrm{d}Y_j.
\end{equation}
Here $X_i\in \Gamma_i\times(-\eta_i,\eta_i)$ for some $i=1,2,\ldots,K$ and $w_{i,j}$ has been introduced to incorporate reflection/transmission coefficients.  An exponential damping term with coefficient $\mu_j> 0$ has also been introduced,  where $D(X_i,Y_j)$ denotes the distance between $s_j$ and the solution point.  The global boundary operator $\mathcal{B}$ is then given by $\mathcal{B}=\sum_j\mathcal{B}_j$, where the sum is taken over each $\Omega_j$ that shares an edge with $\Omega_i$,  including $\Omega_i$ itself. 

The stationary boundary density $\rho$ can be expressed in terms of a Neumann series
\begin{equation}\label{eq:RhoFinal}
\rho =
\sum_{n=0}^{\infty}\mathcal{B}^{n}[\rho_{0}] =
(I-\mathcal{B})^{-1}[\rho_{0}],
\end{equation}
where $\rho_0$ is a given initial boundary density and $\mathcal{B}^{n}$ represents $n$
iterates of the operator $\mathcal{B}$.  Once 
$\rho$ has been evaluated using (\ref{eq:RhoFinal}),  the interior density $\rho_{\Omega}$ can be calculated by projecting onto a prescribed solution point $\mathbf{r}\in\Omega_j$ using \cite{TH19}
\begin{equation}\label{eq:RhoOmega}
\rho_{\Omega}(\mathbf{r}) = \eta_j^2 \int_{0}^{2\pi}
e^{-\mu_j D(\mathbf{r},s_j)}\rho(s_j(\mathbf{r},\Theta),p_j(\mathbf{r},\Theta))\,\mathrm{d}\Theta.
\end{equation}
Here,  $\Theta\in[0,2\pi)$ is the polar angle parametrising trajectories approaching $\mathbf{r}$ from $s_j(\mathbf{r},\Theta)\in \Gamma_j$ and $D$ is used to represent the length of the trajectory between $\mathbf{r}\in \Omega_j$ and $s_j\in\Gamma_j$.

\section{PETROV-GALERKIN DISCRETIZATION}\label{sec:FiniteDimAppr}
\begin{figure}[h]
 \centerline{\includegraphics[trim=0 50 0 10, width=250pt]{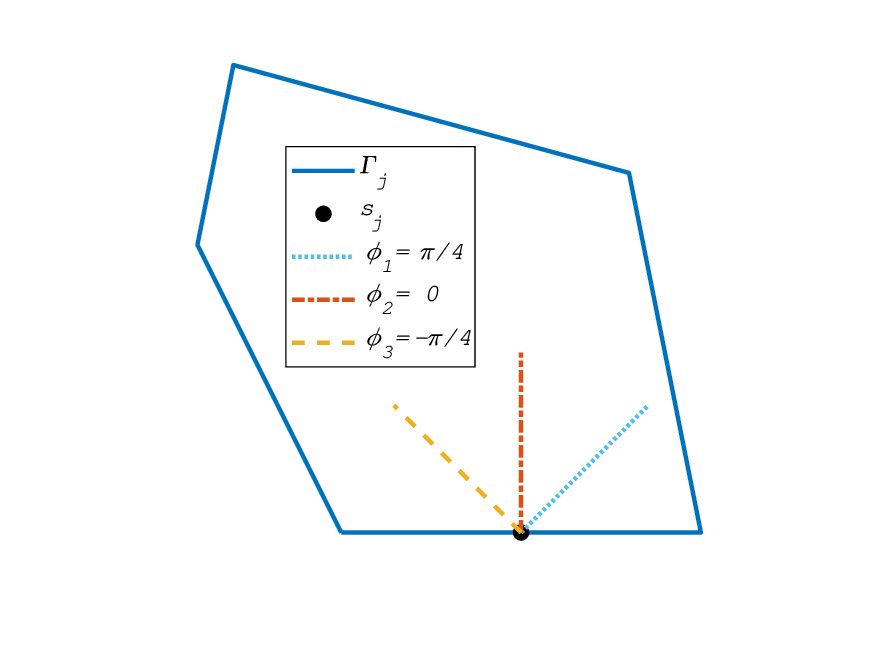}}
\caption{The local direction coordinates $\phi_n(s_j)\in(-\frac{\pi}{2},\frac{\pi}{2})$,  $n=1,2,3$  for $L=8$ global directions given by
 $\Phi_l=2\pi (l-1)/8$ for $l=1,2,\ldots, 8$.  The local and global directions are connected via $\phi_n = \gamma - \Phi_{n+1}$ for $n=1,2,3$,  where $\gamma\in[0,2\pi)$ is the global direction of the inward normal vector at $s_j$ (here $\gamma=\pi/2$).
}\label{fig:LocDir}
\end{figure}

In this section we introduce two direction preserving discretizations of the boundary operator (\ref{eq:OpB}) using Petrov-Galerkin projections in order to numerically solve for $\rho$ by solving a discretised form of equation (\ref{eq:RhoFinal}).  We first split $\Gamma_j$ into boundary elements $E^j_m$ for $m=1,2,\ldots,M_j$ and define the global ray directions $\Phi_l\in[0,2\pi)$ for $l=1,2,\ldots L$.  These global direction are defined anti-clockwise relative to the positive $x_1$-axis as specified in \cite{DeltaDEA}.  Here we take $\Phi_l=2\pi (l-1)/L$,  but note that this choice is flexible and can easily be amended to include dominant transmission paths, where known.  Let $\phi_n(s_j)\in(-\pi/2,\pi/2)$, $n=1,2,\ldots,N_m$ denote the local ray directions at $s_j\in\Gamma_j$.  The local directions at $s_j$ are simply a subset of the global directions taken as all directions which satisfy the property of being directed into $\Omega_j$ from $s_j$.  The local directions are also given a local numbering based on their direction relative to the interior normal at $s_j$ - see Fig.~\ref{fig:LocDir}. 

We apply an approximation of the form
\begin{equation}\label{eq:Expansion}
\rho(s_j,p_j)\approx\sum_{m=1}^{M_j}\sum_{n=1}^{N_m}\rho_{(j,m,n)}
b_{m}(s_j) \delta(p_j-\tilde{p}_n(s_j)), \qquad j=1,\dots,K,
\end{equation}
where $\tilde{p}_n(s_j)=\eta_j\sin(\phi_n(s_j))$ and 
$b_m(s_j)=|E^j_m|^{-1/2}:=(\mathrm{diam}({E^j}_m))^{-1/2}$ for $s_j\in {E^j}_m$ or $b_m(s_j)=0$ otherwise.  We impose a standard (Bubnov) Galerkin projection in $s_j$ with the orthonormal basis $b_m$, $m=1,2,\ldots, M_j$.  It is therefore only in the momentum variable $p_j$ that we apply a Petrov-Galerkin projection and compare two possible choices of test functions that are orthonormal in the $L^2$ inner product to $\delta(p_j-\tilde{p}_n(s_j))$ for $n=1,2,\ldots,N_m$. The first set of test functions that we consider are the indicator functions 
$$\chi_n(p_j)=\tilde\chi_n(\arcsin(p_j / \eta_{j}))$$ 
introduced in \cite{DeltaDEA},  where $\tilde\chi_n(\theta_j)=1$ if $\theta_j\in((\phi_{n-1}+\phi_n)/2,(\phi_n+\phi_{n+1})/2)$ and zero otherwise.
The second set of test functions that we propose are the piecewise linear hat functions $B_n(p_j)=\tilde{B}_n(\arcsin(p_j / \eta_{j}))$,
where
\[\tilde{B}_n(\theta_j)=\left\{\begin{array}{ccc}\displaystyle 1+\frac{\theta_j-\phi_n}{\phi_n-\phi_{n-1}} & \mathrm{if} & \theta_j\in(\phi_{n-1},\phi_n],\vspace{2mm}\\ \displaystyle 1+\frac{\phi_n-\theta_j}{\phi_{n+1}-\phi_{n}} & \mathrm{if} &  \theta_j\in(\phi_n,\phi_{n+1}],\vspace{2mm}\\ 0 &  & \mathrm{otherwise.} \end{array}\right.\]
In both cases we note that
\begin{equation}\label{eq:PetOrth}
\langle\delta(\cdot -\tilde{p}_n(s_j)),\chi_{n'}\rangle_{L^2(-\eta_j,\eta_j)}=\langle\delta(\cdot -\tilde{p}_n(s_j)),B_{n'}\rangle_{L^2(-\eta_j,\eta_j)}=0
\end{equation}
for all $n\neq n'$,  and if $n=n'$ then the right hand side of (\ref{eq:PetOrth}) is instead one.

We now apply the various Galerkin projections described above to the operator $\mathcal{B}$.  Using piecewise linear test functions in direction leads to a matrix representation $B$ of $\mathcal{B}$ as follows:
\begin{eqnarray}\label{eq:TransferBij}
B_{I,J}&=&\int_{\Gamma_j\times(-\eta_j,\eta_j)}\hspace{-9mm}e^{-\mu_j D(\varphi_{i,j}(Y_j),Y_j)} w_{i,j}(Y_j)b_{m}(s_j)b_{m'}(s'_i(Y_j))
\delta(p_j-\tilde{p}_n(s_j))B_{n'}(p'_i(Y_j))\,\mathrm{d}Y_j\vspace{2mm}\nonumber\\
&=&\int_{\Gamma_j}\hspace{-1mm} e^{-\mu_j D_i(s_j)} w_{i,j}(s_j,\tilde{p}_n(s_j))b_{m}(s_j)b_{m'}(s'_i(s_{j},\tilde{p}_n(s_j)))
B_{n'}(p'_i(s_{j},\tilde{p}_n(s_j)))\,\mathrm{d}s_j\\
&=&\frac{w_{i,j}(\tilde{p}_n)}{|E^j_m|^{1/2}}\int_{E^j_m}\hspace{-1mm}e^{-\mu_j D_i(s_j)} b_{m'}(s'_i(s_{j},\tilde{p}_n(s_j)))
B_{n'}(p'_i(s_{j},\tilde{p}_n(s_j)))\,\mathrm{d}s_j,\nonumber
\end{eqnarray}
where  $I=(i,m',n')$ and $J=(j,m,n)$ are multi-indices. To implement the piecewise constant test functions $\chi_{n'}$ as in \cite{DeltaDEA} instead, we simply exchange $B_{n'}$ for $\chi_{n'}$ in (\ref{eq:TransferBij}).  In addition,  the notation $D_i(s_j)$ represents the length of the trajectory between $s_j\in\Gamma_j$ and $s_i'(s_j,\tilde{p}_n(s_j))\in\Gamma_i$.  We remark that the four integrals required for the phase-space Galerkin projection have been simplified to just one integral over $E^j_m\subset\Gamma_j$.  Note that in the third line of (\ref{eq:TransferBij}) we have made the additional assumption that the reflection/transmission coefficients  $w_{i,j}(s_j,\tilde{p}_n(s_j))=w_{i,j}(\tilde{p}_n)$  depend only on the arrival direction of the ray along an interface and not on the position along the interface.  For homogeneous polygonal sub-domains,  the Euclidean distance function $D_i(s_j)$ is  linear in $s_j\in E^j_m$ and hence the one integral remaining in the third line of (\ref{eq:TransferBij}) can be evaluated relatively simply.  One can also apply analytic spatial integration for higher order (Legendre polynomial) spatial basis functions as detailed in \cite{JB17}.

The expansion coefficients in (\ref{eq:Expansion}) can now be obtained by solving a linear system
$${\bm{\rho}} = (I-B)^{-1}{\bm{\rho}}_0$$
corresponding to the discretized operator equation (\ref{eq:RhoFinal}).  Here, the vectors $\bm{\rho}_0$ and
$\bm{\rho}$ contain the expansion coefficients for $\rho_{0}$ and $\rho$,  respectively.  The  source vector
$\bm{\rho}_0$ can be evaluated from the prescribed initial density $\rho_0$ by making use of the orthonormality properties of the bases in both position and momentum as follows
$$
[\bm{\rho}_0]_J=\int_{\Gamma_j\times(-\eta_j,\eta_j)}\rho_0(s_j,p_j)b_m(s_j)B_{n}(p_j)\mathrm{d}Y_j=\frac{\eta_j}{|E^j_m|^{1/2}}\int_{E^j_m}\int_{\phi_{n-1}}^{\phi_{n+1}}\rho_0(s_j,p_j(\theta_j))\tilde{B}_n(\theta_j)\cos(\theta_j)\mathrm{d}\theta_j\mathrm{d}s_j.
$$
The corresponding expression when using the test functions $\chi_n$ instead of $B_n$ is slightly simpler since the piecewise constants are independent of the integration variable and hence
$$
[\bm{\rho}_0]_J=\int_{\Gamma_j\times(-\eta_j,\eta_j)}\rho_0(s_j,p_j)b_m(s_j)\chi_{n}(p_j)\mathrm{d}Y_j=\frac{\eta_j}{|E^j_m|^{1/2}}\int_{E^j_m}\int_{(\phi_{n-1}+\phi_n)/2}^{(\phi_n+\phi_{n+1})/2}\rho_0(s_j,p_j(\theta_j))\cos(\theta_j)\mathrm{d}\theta_j\mathrm{d}s_j.
$$
Numerical results comparing these discretization schemes will be presented during the conference and included in an extended version of this paper.
% Acknowledgement
\section{ACKNOWLEDGMENTS}
Support from the EPSRC (grant no.~EP/R012008/1) is
gratefully acknowledged. 

% References

%\nocite{*}
\bibliographystyle{aipnum-cp}%

\end{document}